\documentclass[epj,a4paper]{svjour}
\usepackage{graphics}
\def\lsim{\raise0.3ex\hbox{$<$\kern-0.75em\raise-1.1ex\hbox{$\sim$}}}
\def\gsim{\raise0.3ex\hbox{$>$\kern-0.75em\raise-1.1ex\hbox{$\sim$}}}
\begin{document}
\title{Lattice QCD thermodynamic results with improved staggered fermions}
\author{Christian Schmidt\inst{1} for RBC-Bielefeld and HotQCD Collaborations
}                     % Do not remove
\titlerunning{Lattice QCD thermodynamic results}
\authorrunning{C. Schmidt}
\institute{Fakult\"at f\"ur Physik, Universit\"at Bielefeld, 
D-33615 Bielefeld, Germany}
\date{Received: date / Revised version: date}
\abstract{
We present results on the QCD equation of state, obtained with two
different improved dynamical staggered fermion actions and almost
physical quark masses. Lattice cut-off effect are discussed in detail
as results for three different lattice spacings are available now,
i.e. results have been obtained on lattices with temporal extent of
$N_\tau=4,6$ and 8. Furthermore we discuss the Taylor expansion
approach to non-zero baryon chemical potential and present 
the isentropic equation of state on lines of constant entropy per baryon
number.
\PACS{
      {11.15.Ha}{Lattice gauge theory}   \and
      {11.10.Wx}{Finite-temperature field theory} \and
      {12.38.Gc}{Lattice QCD calculations} \and
      {12.38.Mh}{Quark-gluon plasma}
     } % end of PACS codes
} %end of abstract
\maketitle
\section{Introduction}
\label{intro}
A detailed and comprehensive understanding of the thermodynamics
of quarks and gluons, e.g. of the equation of state is most desirable
and of particular importance for the phenomenology of relativistic
heavy ion collisions.
Lattice regularized QCD simulations at non-zero temperatures have
been shown to be a very successful tool in analyzing the
non-perturbative features of the quark-gluon plasma. Driven by both,
the exponential growth of the computational power of recent
super-computer as well as by drastic algorithmic improvements one is
now able to simulate dynamical quarks and gluons on fine lattices with
almost physical masses. 

In this article we present results on bulk thermodynamic quantities
and the equation of state on lattices with temporal extent of
$N_\tau=4,6$ \cite{EoS} and preliminary results on $N_\tau=8$,
obtained by the HotQCD collaboration\cite{hotQCDeos,hottalks}. 
The article is
organized as follows, in Sec.~\ref{action} we discuss details of our
lattice actions and corresponding finite cut-off corrections, in
Sec~\ref{lcp} we present our choice of lattice parameter, in
Sec.~\ref{eos} we present preliminary results on the equation of state
and in Sec.~\ref{density} we introduce the Taylor expansion method and
calculate leading and next to leading order corrections of bulk
thermodynamic quantities to a non-zero baryon chemical
potential. Finally, we present our preliminary results on the
isentropic equation of state in Sec.~\ref{sec:isen} and conclude in
Sec.~\ref{sec:con}.

\section{Choice of action and cut-off effects}
\label{action}
In order to control lattice cut-off effects it is of particular
importance for finite temperature calculations to improve the lattice
action beyond the naive discretization scheme. As most bulk thermodynamic
observables, like the energy density and pressure, are dimension four
operators, the numerical signal for these observables drops like the
lattice spacing to the fourth power. One is forced to perform 
lattice calculations on rather coarse lattices, where cut-off effects
are still sizable. Using an improvement scheme as introduced by
Symanzik \cite{Symanzik} for the gauge part of the action and
similarly for the fermionic part in the case of staggered fermions
\cite{Naik,p4} or Wilson type fermions
\cite{clover,perfect1,perfect2}, cut-off effects can be drastically
reduced. By adding irrelevant operators to the action which will
vanish in the continuum limit, cut-off effect to arbitrary order in
$\mathcal{O}(a^n)$ can be eliminated already for finite lattice spacing
$a$. The above mentioned improvement schemes for staggered fermions are
tree-level improvements $\mathcal{O}(g^0)$ but can in principle be
generalized to eliminate cut-off effect also in leading order of the
gauge coupling $\mathcal{O}(g^2)$.

The basis for the two actions we will consider here is given by
two versions of $\mathcal{O}(a^2g^0)$ improved staggered fermions which
contain quark and anti-quark fields separated by up to three links. By
adding a straight 3-link term to the standard 1-link term, and
choosing the coefficient such that the leading order cut-off effects
cancel, one arrives at the Naik action \cite{Naik}. Using bended
3-link terms instead (knight moves), one derives the p4 action
\cite{p4}. The latter action has a dispersion relation which is
$\mathcal{O}(p^4)$ improved.

In addition to the tree-level improvements one introduces a smeared
1-link term which reduces the flavor symmetry breaking of the staggered
fermions. In case of the p4-action, the 1-link term is smeared by the
sum of all corresponding 3-link staples. In case of the Naik action,
staples up to length of 7 links are used as well as an additional
tadpole improvement (1-loop level). The gauge part of both actions is
Symanzik improved. Using all this improvements together these
actions are called p4fat3 and asqtad, respectively.

\begin{figure}
\begin{center}
\resizebox{0.45\textwidth}{!}{%
  \includegraphics{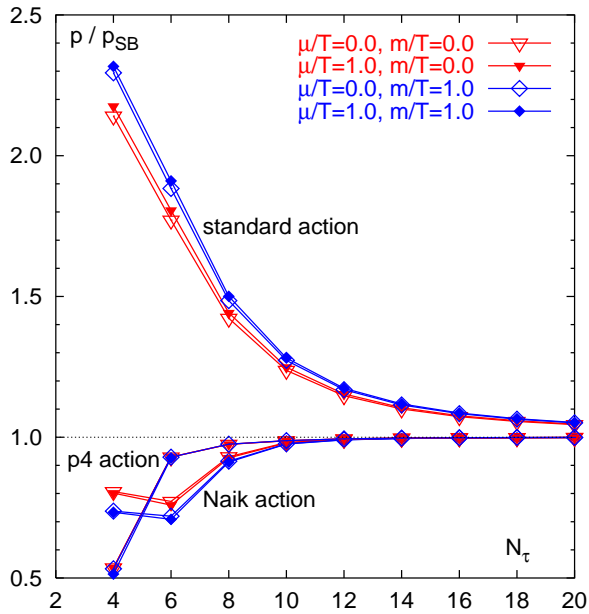}
}
\end{center}
\caption{Pressure for three different types of free lattice fermions
as function of $N_\tau$. Results are given in units of the continuum
ideal Fermi gas value and are calculated for two different masses and
two different values of chemical potentials.}
\label{fig:cutoff}  
\end{figure}

In the free gas limit ($T\to\infty$), where the smearing becomes
irrelevant, the cut-off effects of the p4 and Naik actions for
bulk thermodynamic quantities have been analyzed in detail
\cite{Hegde}. Within a large $N_\tau$ expansion, where $N_\tau$ is the
number of lattice sites in temporal direction one can quantify how the
pressure approaches its continuum (Stefan-Boltzmann) value $p_{SB}$. In
comparison we find for the standard (non-improved), p4 and the Naik
actions
\begin{equation}
\begin{array}{cclc}
  \frac{p}{p_{SB}}\!=\!1\! + \!\!&
  \frac{248}{147}\! \left(\frac{\pi}{N_\tau}\right)^2 \!\!\!\! &
+ \frac{635}{147}\! \left(\frac{\pi}{N_\tau}\right)^4 + \cdots &
\mbox{(std)} \\
  \frac{p}{p_{SB}}\!=\!1\! + \!\!&
 0 &
- \frac{1143}{980}\!\left(\frac{\pi}{N_\tau}\right)^4\!\!\!\!
+ \frac{ 73}{2079}\!\left(\frac{\pi}{N_\tau}\right)^6\!\!\!\! +\! \cdots\!\!\!\! &
\mbox{(p4)} \\
  \frac{p}{p_{SB}}\!=\!1\! + \!\!&
 0 & 
- \frac{1143}{980}\!\left(\frac{\pi}{N_\tau}\right)^4\!\!\!\!
- \frac{365}{77}\!\left(\frac{\pi}{N_\tau}\right)^6\!\!\!\! +\! \cdots\!\!\!\! &
\mbox{(Naik)}
\end{array}
\end{equation}
As one can see, the leading order corrections of the improved p4 and
Naik actions are identical. However, the sub-leading coefficient which
is of order $(\pi/N_\tau)^6$ is much smaller for p4 than for the Naik
action. In Fig~\ref{fig:cutoff} we show the exact evaluation of the
pressure as function of $N_\tau$, for two different masses and two
different chemical potentials.
It is evident, that the dependence of cut-off effects on quark mass
and chemical potential $\mu$ is very small and similar in all three
discretization schemes. In fact, one can show that the dependence of
finite cut-off corrections on $\mu$ is given by Bernoulli polynomials
and is independent of the discretization schemes \cite{Hegde}.

\section{Lattice parameter and scale setting}
\label{lcp}
We perform calculations on lattices of extent $16^3\times 4$,
$24^3\times 6$, using the p4fat3 action \cite{EoS} and $32^3\times 8$
with both p4fat3 and asqtad actions. The latter calculations are still 
preliminary and are currently performed
by the HotQCD collaboration \cite{hotQCDeos}.  For the generation of
gauge configurations we use the exact RHMC algorithm \cite{RHMC}.  For
each finite temperature calculation we perform a corresponding zero
temperature calculation on a lattice of at least the size
$N_\sigma^4$, where $N_\sigma$ is the spatial extent of the finite
temperature calculation.

The simulations are done on a line of constant physics (LCP), i.e.
the quark masses are kept constant in physical units. In practice, this has been
obtained by tuning the bare quark masses such that the meson masses of e.g.
pion, kaon and pseudo-scalar strange meson $\bar ss$ stay
constant in the QCD vacuum as we change the value of the coupling.
The strange quark mass was always fixed to its
physical value, by fixing kaon and $\bar s s$ to their
corresponding physical values \cite{EoS}. We find that the LCP can, to a good
approximation, be parameterized by a constant ratio of the bare
quark masses. Most calculation are done on a LCP which corresponds to
a pion mass of about 220 MeV. We do, however, also show preliminary
results with a physical pion mass, i.e. $m_\pi\approx 150$ MeV.

To set set the temperature scale in physical units, we determine two
distance scales, $r_0$ and $r_1$, from the zero temperature static
quark potential
\begin{equation}
\left(r^2\frac{{\rm d}V_{\bar q q}(r)}{{\rm d}r}\right)_{r=r_0}=1.65,\;
\left(r^2\frac{{\rm d}V_{\bar q q}(r)}{{\rm d}r}\right)_{r=r_1}=1.0
\end{equation}
The ratio of both scales is only slightly quark mass dependent. It has
been determined in both discretization schemes consistently,
$r_0/r_1=1.4636(60)$ (p4fat3 \cite{EoS}) and 1.474(7)(18) (asqtad
\cite{asqtad}). The distance scales $r_0$ and $r_1$ have been related
to properties of the chamonium spectrum which allows to determine them
in physical units. We use here $r_0=0.469(7)$ fm as determined
in Ref. \cite{r0}. More details on the scale setting procedure, as well
as the parameterization of the LCP are given in Ref. \cite{EoS}.

\section{The equation of state}
\label{eos}
\begin{figure*}
\begin{center}
\resizebox{0.575\textwidth}{!}{%
  \includegraphics{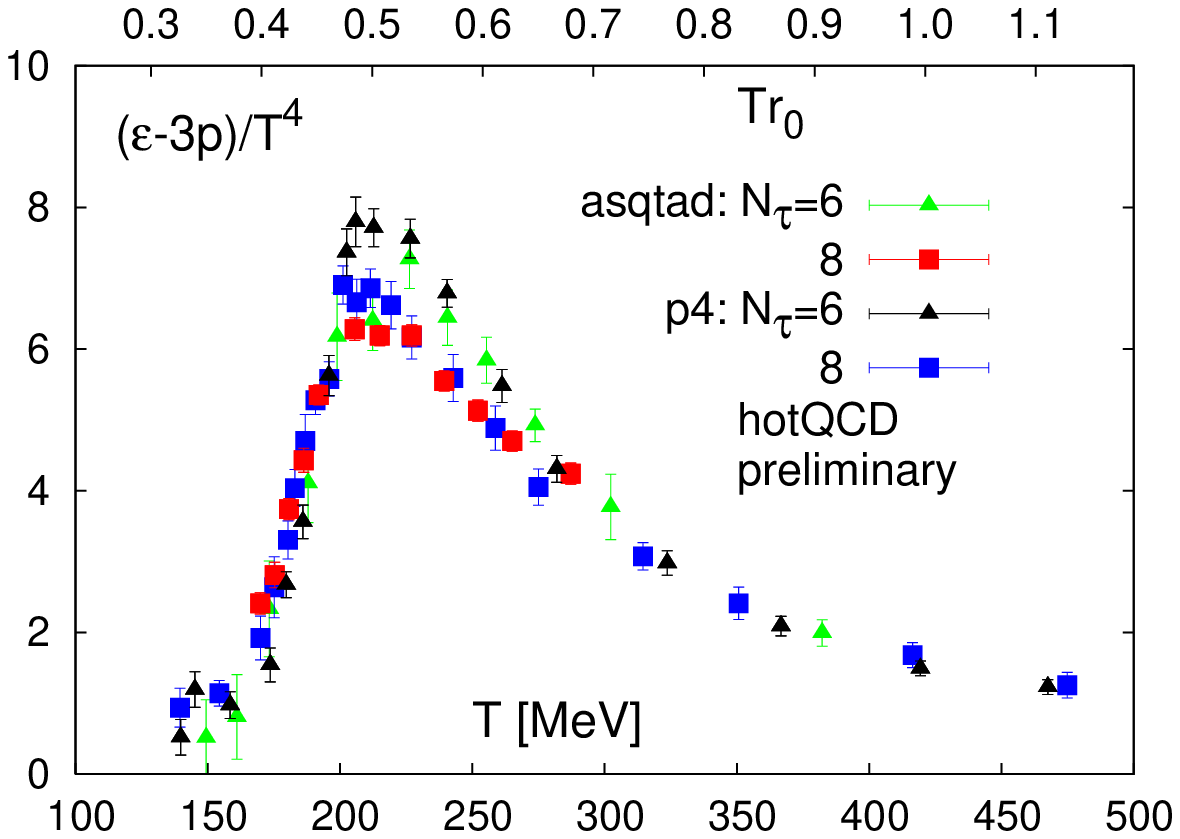}
}
\resizebox{0.4\textwidth}{!}{%
  \includegraphics{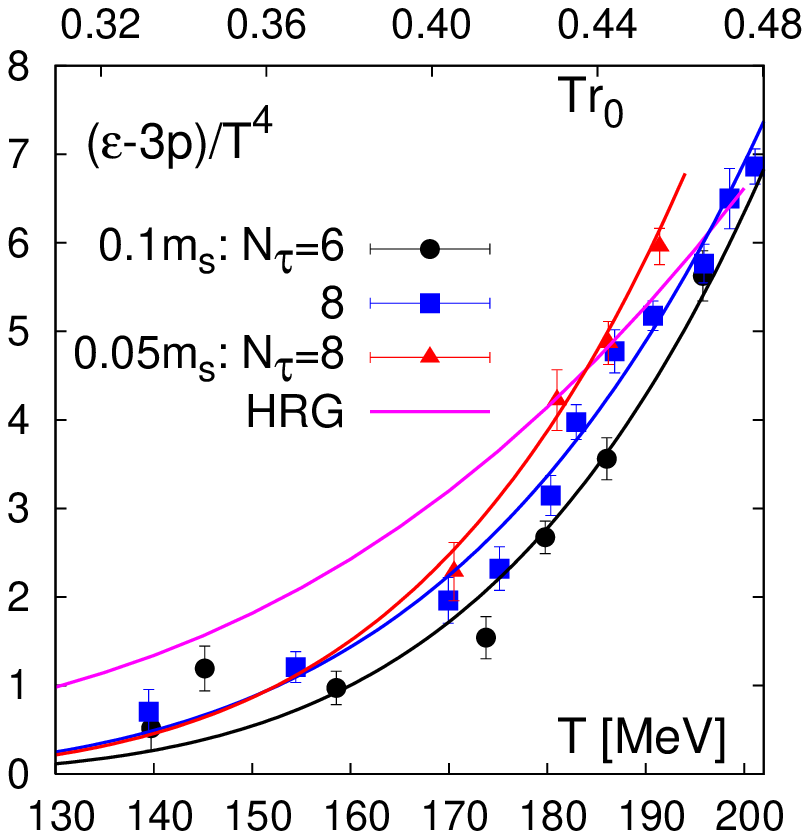}
}
\end{center}
\caption{ 
The trace anomaly, $(\epsilon -3p)/T^4$ on $N_\tau=6$ and $8$
lattices. On the left panel results are obtained with the p4fat3 and
asqtad actions, where $N_\tau=6$ results are from \cite{EoS} and
\cite{milc_eos} respectively, $N_\tau=8$ results are preliminary
hotQCD results \cite{hotQCDeos}. The right panel shows the low
temperature part of the trace anomaly in detail, here we plot only
results obtained by the p4fat3 action. Also shown on the right panel
are quadratic fits to the data, as well as the trace anomaly obtained
in the framework of the Resonance gas model. Light quark masses have
been constrained to be one tenth of the strange quark mass ($0.1m_s$), 
on the right panel we show, however, also preliminary results from
simulations with physical quark masses ($0.05m_s$).}
\label{fig:e-3p}  
\end{figure*}

Along the line of constant physics, at sufficiently large volume and at zero chemical
potential, the temperature is the only intensive parameter that
controls the thermodynamics.  Consequently there exists only one
independent bulk thermodynamic observable that needs to be
calculated. All other quantities are than obtained by using standard
thermodynamic relations. On the lattice, it is convenient to first
calculate the trace anomaly in units of the fourth power of the
temperature, $\Theta^{\mu\mu}/T^4$. It is easily obtained as a
derivative of the pressure $p/T^4$, with respect to the temperature,
\begin{equation}
\frac{\Theta^{\mu\mu}}{T^4} \equiv \frac{\epsilon-3p}{T^4}
=T\frac{\partial}{\partial T}(p/T^4).
\label{eq:trace}
\end{equation} As the pressure is directly given by the partition
function, $p/T=V^{-1}\ln Z$, the calculation of the trace anomaly
requires only the evaluation of rather simple expectation
values. According to Eq. \ref{eq:trace} one then obtains the pressure
by
\begin{equation} \frac{p(T)}{T^4} - \frac{p(T_0)}{T_0^4} =
\int_{T_0}^{T} {\rm d}T' \frac{1}{T'^5} \Theta^{\mu\mu} (T') \;\; .
\label{pres}
\end{equation} Here $T_0$ is an arbitrary temperature value which
usually is chosen in the low temperature regime where the pressure and
other thermodynamic quantities are suppressed exponentially by
Boltzmann factors corresponding to the lightest hadronic states;
e.g. the pions.  The energy density is then obtained by combining
results for $p/T^4$ and $(\epsilon -3p)/T^4$, respectively.

In Fig.~\ref{fig:e-3p} (left) we show results for $\Theta^{\mu\mu}/T^4$
obtained with the asqtad and p4fat3 actions, respectively. The new
$N_\tau=8$ results \cite{hotQCDeos} are compared to $N_\tau=6$ results
taken from \cite{EoS,milc_eos}.  We note that the asqtad and p4fat3
formulations give results which are in good agreement with each
other. In fact, in quite a large temperature regime the agreement for
given lattice extent $N_\tau$ seems to be better than one could expect
in view of the overall cut-off dependence that is visible when
comparing results for $N_\tau=6$ and $N_\tau=8$ more closely.
They lead to a reduction of the peak height in $\Theta^{\mu\mu}/T^4$,
which is located at $T\simeq 200$~MeV and to a shift of the 
rapidly rising part of $\Theta^{\mu\mu}/T^4$ in the transition region to
smaller values of the temperature.  

In Fig.~\ref{fig:e-3p} (right) we show quadratic fits to
the data obtained by the p4fat3 action, to highlight the cut-off
effects. It is evident, that the $N_\tau=8$ data is shifted relative
to the $N_\tau=6$ data by about $8$~MeV at low temperatures, $T\simeq
160$~MeV. This shift decreases to about $5$~MeV at temperatures
$T\simeq 190$~MeV.

Light quark masses have been constrained to be one tenth of the
strange quark mass ($0.1m_s$), on the right panel we show, however,
also preliminary results from simulations with physical quark masses
($0.05m_s$). This again leads to a further shift in $\Theta^{\mu\mu}$
of approximately $5$~MeV at $T\simeq 190$~MeV towards lower
temperatures.

We also compare the results for $(\epsilon -3p)/T^4$ to results
obtained from the hadron resonance gas model. Details on the resonance
gas curve in Fig.~\ref{fig:e-3p} (right) will be given in
\cite{hotQCDeos}. The slope of $(\epsilon -3p)/T^4$ obtained by the
resonance gas model seems to be much smaller than the slope obtained
by the quadratic fits to the data.  Whether this points at deviations
of the equation of state at lower temperatures from resonance gas
behavior or is due to larger cut-off effects in the low temperature
regime requires further studies. We note that the lattice spacing
becomes larger at lower temperatures and violations of flavor
symmetry, which are inherent to the staggered fermion formulations at
finite lattice spacing, thus may become more important.
%Thus the preliminary data suggests
%that there exist only a relative small temperature interval of
%$175$~MeV~$\lsim T \lsim$~$195$~MeV, where one can expect that the
%resonance gas picture gives a reasonable good description of QCD.
%However, the low temperature regime of $T\lsim 180$~MeV clearly 
%requires a more detailed analysis with more data points at $m_l=0.05m_s$.

\begin{figure}
\begin{center}
\resizebox{0.48\textwidth}{!}{%
  \includegraphics{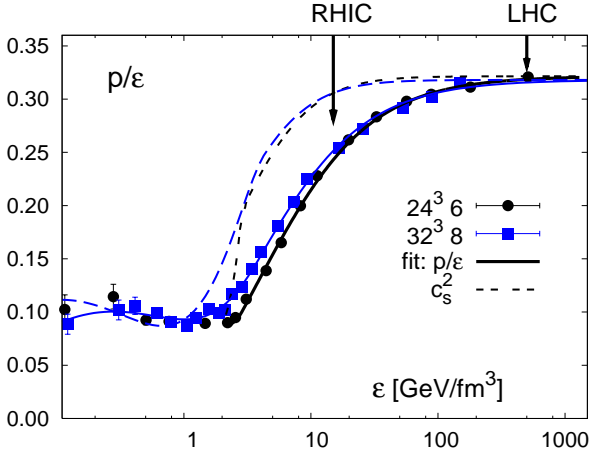}
}
\end{center}
\caption{The ratio of pressure and energy density as well as the
velocity of sound obtained in calculations with the p4fat3 action on
$N_\tau=6$ \cite{EoS} and $N_\tau=8$ \cite{hotQCDeos} lattices.}
\label{fig:eos}  
\end{figure}

The cut-off dependence observed in $\Theta^{\mu\mu}/T^4$ carries over
to the calculation of pressure and energy density; the former is
obtained by integrating over $\Theta^{\mu\mu}/T^5$ and the energy
density is then obtained by combining results for $p/T^4$ and
$(\epsilon -3p)/T^4$.  This is apparent in Fig.~\ref{fig:eos} where we
show the ratio $p/\epsilon$ obtained with the p4fat3 action on
$N_\tau=6$ \cite{EoS} and $N_\tau=8$ \cite{hotQCDeos} lattices.
Cut-off effects are still visible in the vicinity of the 'softest
point' of the EoS, which is related to the peak position of
$(\epsilon-3p)/T^4$. We find that in the entire range of energy densities relevant
for the expansion of dense matter created at RHIC, $\epsilon \;
\lsim\; 10$~GeV/fm$^3$, the ratio $p/\epsilon$ deviates significantly
from the conformal, ideal gas value $p/\epsilon =1/3$.

This also is reflected in the behavior of the velocity of sound,
$c_s^2 = {\rm d}p/{\rm d}\epsilon$, which is shown in
Fig.~\ref{fig:eos} by dashed lines. It starts deviating significantly
from the ideal gas value below $\epsilon \; \simeq\; 10$~GeV/fm$^3$
and reaches a value of about $0.1$ in the transition region at energy
densities $\epsilon\;\simeq\; 1$~GeV/fm$^3$.  Below the transition it
slightly rises again, but note, that for very small temperatures
$c_s^2$, as well as $p/\epsilon$ are sensitive the integration constant
$p_0(T_0)$ (see Eq.~\ref{eq:trace}). At present we have set $p_0(T_0)=0$,
for $T_0=100$~MeV.

\section{Non-zero chemical potential}
\label{density}
\begin{figure*}
\begin{center}
\resizebox{0.32\textwidth}{!}{%
  \includegraphics{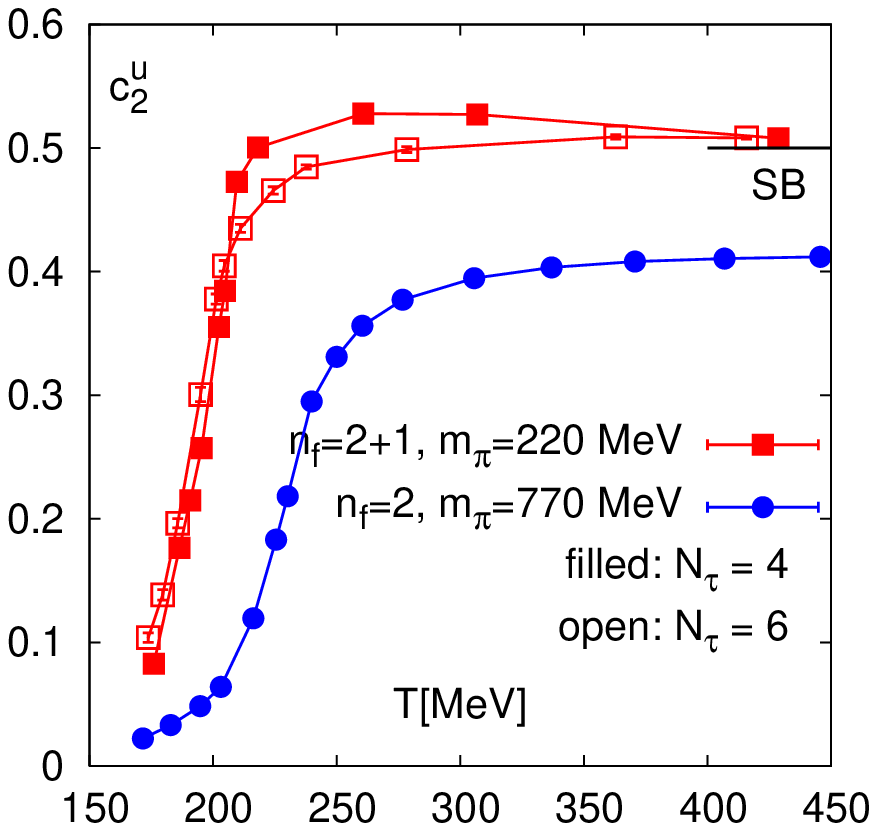}
}
\resizebox{0.32\textwidth}{!}{%
  \includegraphics{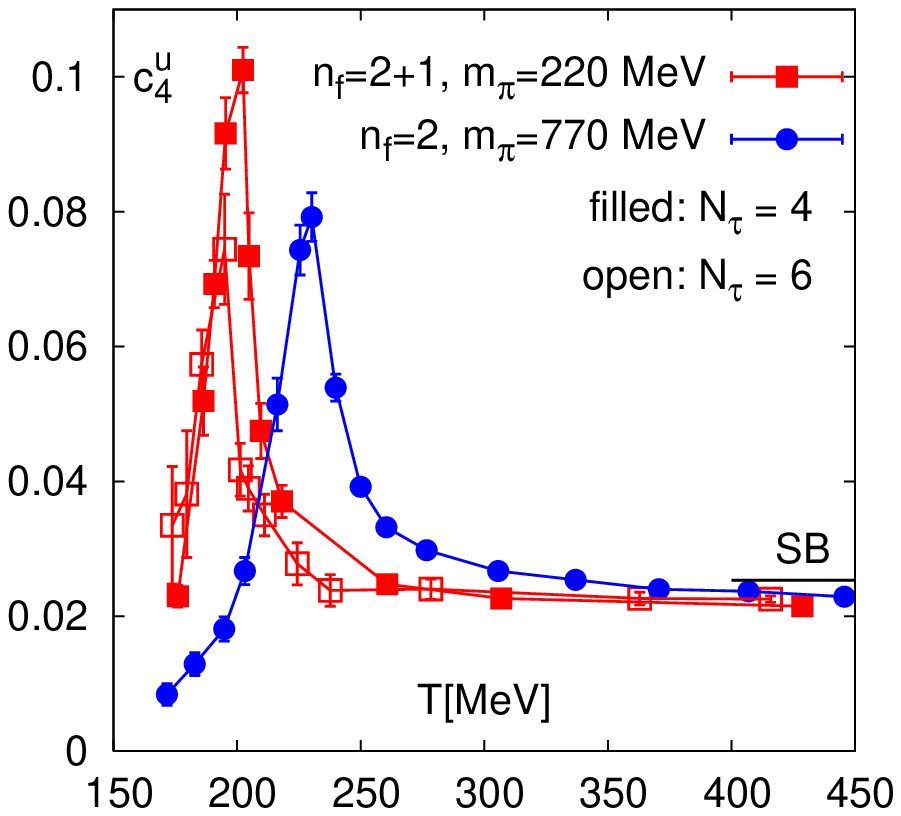}
}
\resizebox{0.32\textwidth}{!}{%
  \includegraphics{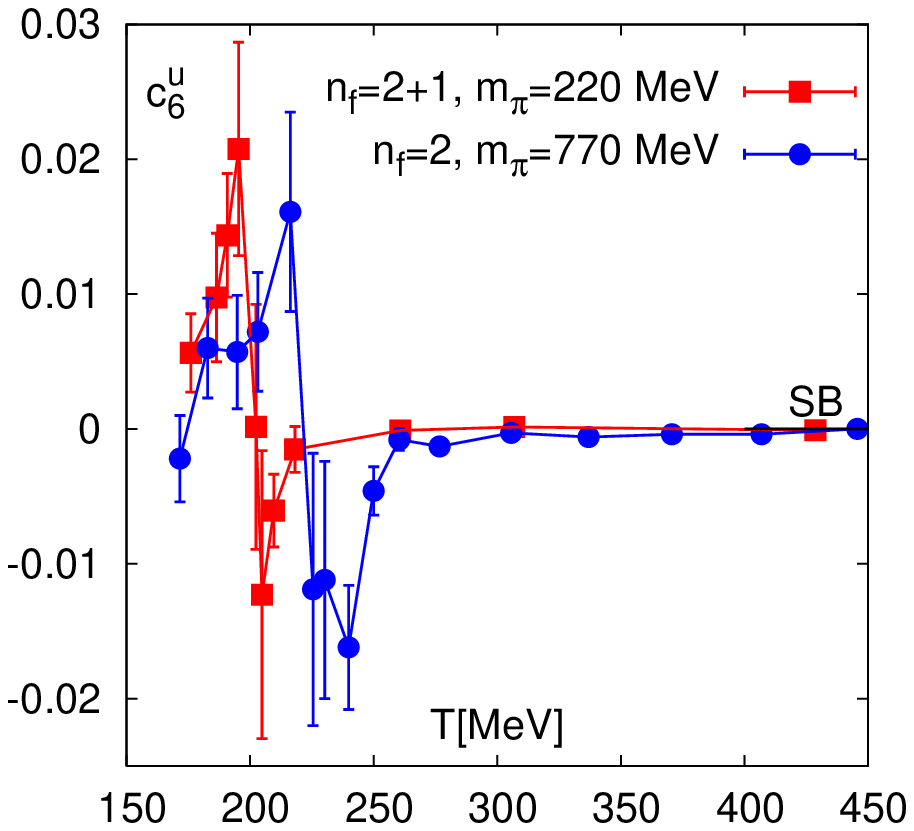}
}
\end{center}
\caption{Taylor coefficients of the pressure in term of the up-quark
chemical potential. Results are obtained with the p4fat3 action on
$N_\tau=4$ (full) and $N_\tau=6$ (open symbols) lattices. We compare
preliminary results of (2+1)-flavor a pion mass of $m_\pi\approx
220$~MeV to previous results of 2-flavor simulations with a
corresponding pion mass of $m_p\approx770$ \cite{eos6}.}
\label{fig:coff_uds}  
\end{figure*}
\begin{figure*}
\begin{center}
\resizebox{0.32\textwidth}{!}{%
  \includegraphics{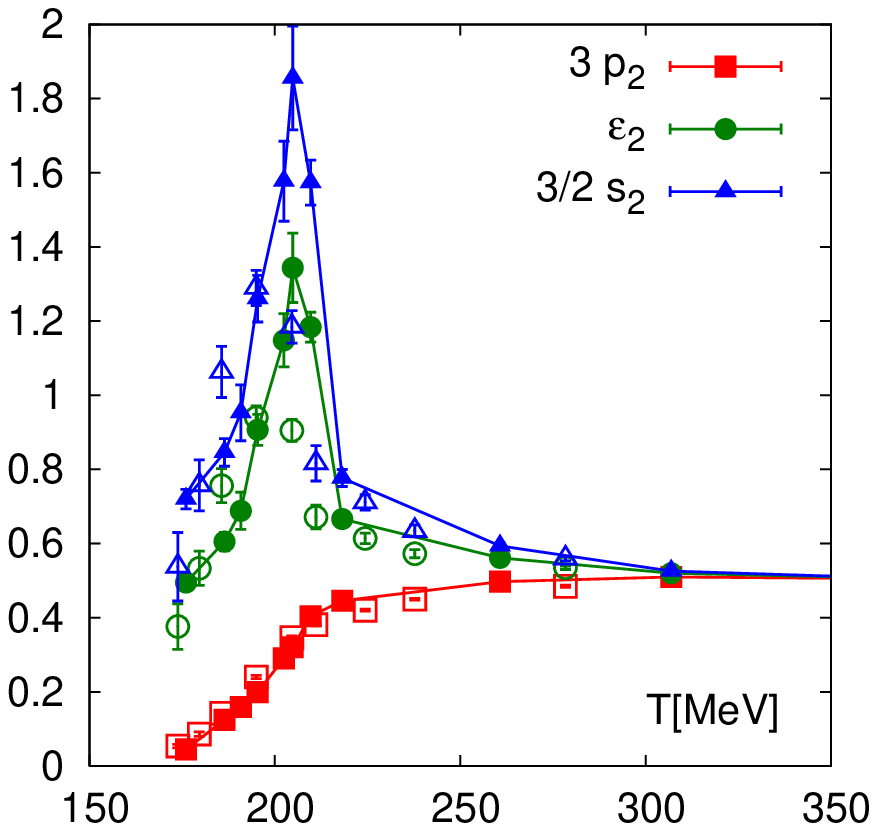}
}
\resizebox{0.32\textwidth}{!}{%
  \includegraphics{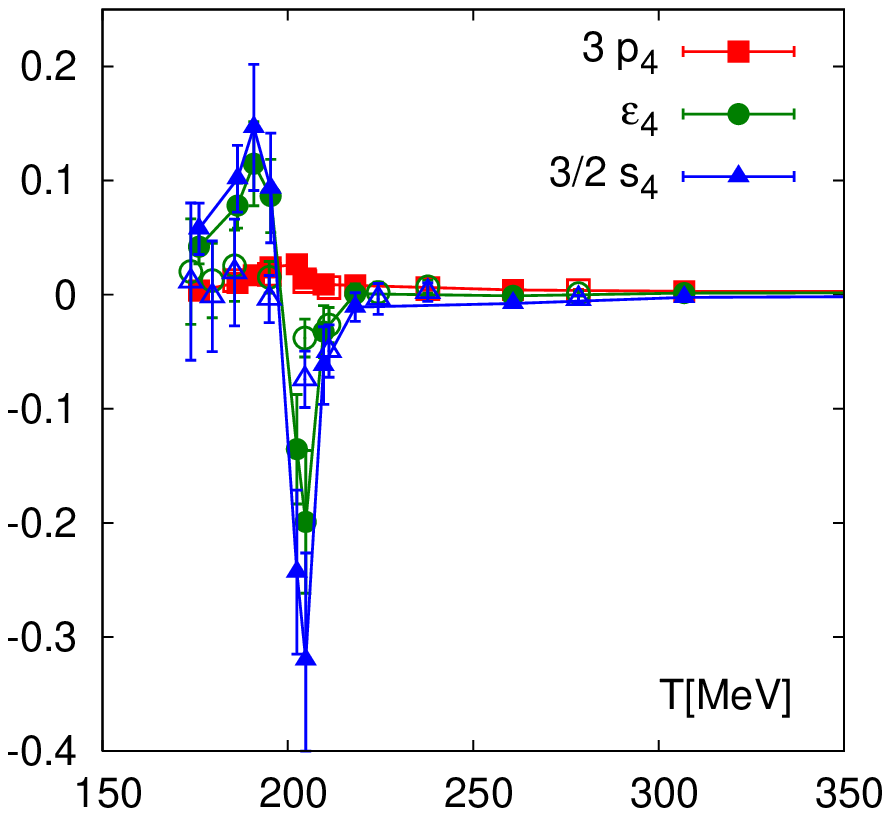}
}
\resizebox{0.32\textwidth}{!}{%
  \includegraphics{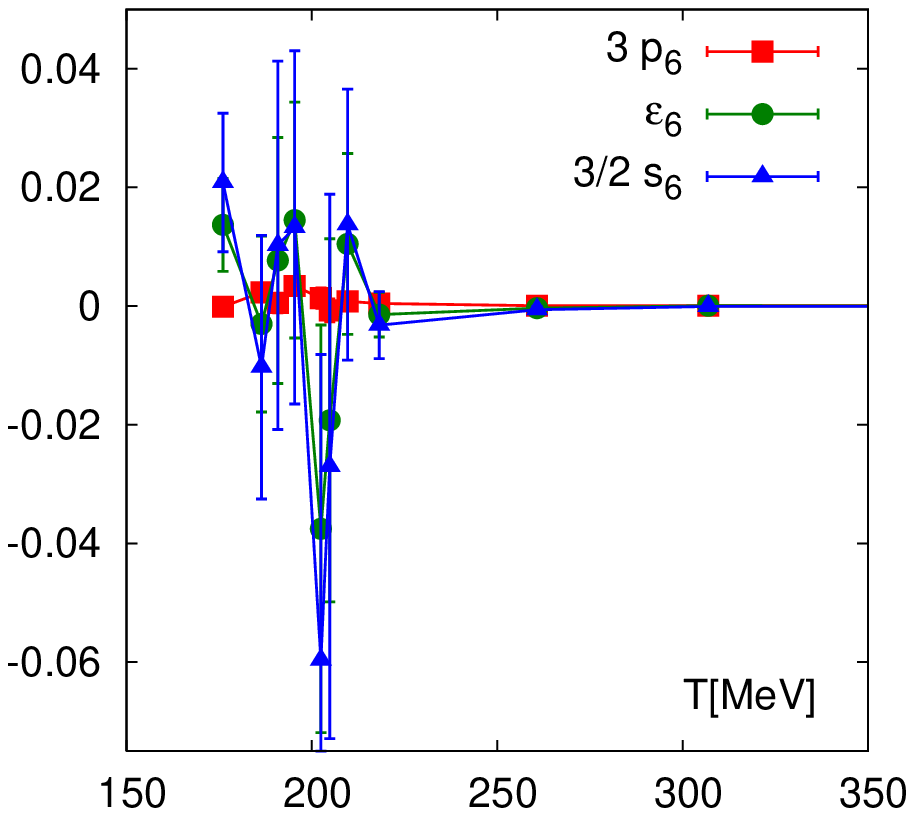}
}
\end{center}
\caption{Taylor coefficients of the pressure, the energy and entropy
density with respect to the baryon chemical potential. Results are
obtained with the p4fat3 action on $N_\tau=4$ (full) and $N_\tau=6$
(open symbols) lattices.}
\label{fig:coff}  
\end{figure*}

At non-zero chemical potential, lattice QCD is harmed by the
``sign-problem'', which makes direct lattice calculations with
standard Monte Carlo techniques at non-zero density practically
impossible.  However, for small values of the chemical potential, some
methods have been successfully used to extract information on the
dependence of thermodynamic quantitites on the chemical potential.
For an overview see, e.g. \cite{overview}.

We closely follow here the approach and notation used in
Ref.~\cite{eos6}. We start with a Taylor expansion for the pressure in
terms of the quark chemical potentials
\begin{equation}
\frac{p}{T^{4}}
=\sum_{i,j,k}c^{u,d,s}_{i,j,k}(T)\left(\frac{\mu_{u}}{T}\right)^{i}
\left(\frac{\mu_{d}}{T}\right)^{j}\left(\frac{\mu_{s}}{T}\right)^{k}.
\label{eq:PTaylor}
\end{equation}
The expansion coefficients $c^{u,d,s}_{i,j,k}(T)$ 
are computed on the lattice at zero chemical potential, using  
stochastic estimators. Some details are given in~\cite{details}.

In Fig.~\ref{fig:coff_uds} we show results on the diagonal expansion
coefficients with respect to the up-quark chemical potential up to the
six order ($c_{n,0,0}^{u,d,s}$ with $n=2,4,6$), obtained with the
p4fat3 action. Here the full symbols are from $N_\tau=4$ lattices, while
the open symbols denote results from $N_\tau=6$ lattices. We find that
cut-off effects are small and of similar magnitude as those found for 
the trace anomaly $\Theta^{\mu\mu}$. This was already
anticipated by the analysis of the cut-off corrections in the free gas
limit. Similar results for the asqtad action have been obtained 
in~\cite{milc_dens}.  

We also compare our preliminary results for 2+1-flavor QCD and a pion
mass of $m_\pi\approx 220$~MeV, with previously obtained result of
2-flavor QCD and $m_\pi\approx 770$~MeV \cite{eos6} (also p4fat3 and $N_\tau=4$).
It is apparent from
Fig.~\ref{fig:coff_uds} that the critical temperature for these
two particular sets of lattice parameter differ substantially and
in fact decreases from about $225$~MeV for the heavier mass calculations
to about $200$~MeV for the lighter mass calculations. Note, that those
$T_c$ values are the $N_\tau=4$ values, which of course are still influenced by the 
finite lattice spacing.  Furthermore, we find from
Fig.~\ref{fig:coff_uds} that the quark number fluctuations of second,
fourth and sixth order, which are related to those expansion
coefficients, increase with decreasing quark mass.

Alternatively to the quark chemical potentials one can introduce
chemical potentials for the conserved quantities baryon number $B$,
electric charge $Q$ and strangeness $S$ ($\mu_{B,Q,S}$), which are
related to $\mu_{u,d,s}$ via
\begin{eqnarray}
\mu_{u}&=&\frac{1}{3}\mu_{B} +\frac{2}{3} \mu_{Q},\\
\mu_{d}&=&\frac{1}{3}\mu_{B}-\frac{1}{3}\mu_{Q},\\
\mu_{s}&=&\frac{1}{3}\mu_{B}-\frac{1}{3}\mu_{Q}-\mu_{S}.
\label{eq:chempot}
\end{eqnarray}
By means of these relations the coefficients $c^{B,Q,S}_{i,j,k}$ of
the pressure expansion in terms of $\mu_{B,Q,S}$ are easily obtained,
in analogy to Eq.~\ref{eq:PTaylor}
\begin{equation}
\frac{p}{T^{4}}
=\sum_{i,j,k}c^{B,Q,S}_{i,j,k}(T)\left(\frac{\mu_{B}}{T}\right)^{i}
\left(\frac{\mu_{Q}}{T}\right)^{j}\left(\frac{\mu_{S}}{T}\right)^{k}.
\label{eq:PTaylor_hadronic}
\end{equation}
For the rest of this article we will restrict ourselves to the case of 
$\mu_Q\equiv\mu_S\equiv0$, thus we will suppress in the following the 
indices that are related to those chemical potentials. From the pressure 
we immediately obtain the baryon number density
$n_B$, which is given by the derivative of $p/T^4$ with respect to the
baryon chemical potential $\mu_B$ and can be expressed in term of the
expansion coefficients $c_{n}^{B}$, we have
\begin{equation}
\frac{n_B}{T^{3}}
=\sum_{n=2}^{\infty}nc^{B}_{n}(T)\left(\frac{\mu_{B}}{T}\right)^{n-1}.
\label{eq:nB_hadronic}
\end{equation}

Using standard thermodynamic relations we can also calculate the expansion 
coefficients of the trace anomaly $\Theta^{\mu\mu}$ or equivalently the 
difference between energy density and three times the pressure,
\begin{equation}
\frac{\epsilon - 3p}{T^4} = \sum_{n=0}^\infty \bar{c}_{n}^{B}(T) 
\left(\frac{\mu_B}{T}\right)^B,
\label{eq:e3pTaylor}
\end{equation}
where the expansion coefficients $\bar{c}_{n}^{B}$ are given by
\begin{equation}
\bar{c}_{n}^{B}(T) 
=T \frac{ {\rm d} c_{n}^{B} (T) }{{\rm d} T}.
\label{eq:cne3p}
\end{equation}
Combining Eqs.~\ref{eq:PTaylor_hadronic},~\ref{eq:e3pTaylor}, and 
\ref{eq:cne3p} we then obtain the Taylor expansions for the energy and 
entropy densities \cite{isen_eos}
\begin{eqnarray}
\frac{\epsilon}{T^4} &=& \sum_{n=0}^\infty \left(3 c_n^B(T) +
\bar{c}_n^B(T)\right) \left(\frac{\mu_B}{T}\right)^n \nonumber \\
              &\equiv& \sum_{n=0}^{\infty} \epsilon_n\left(\frac{\mu_B}{T}\right)^n,\\
\frac{s}{T^3} &\equiv& \frac{\epsilon +p-\mu_B n_B}{T^4}\nonumber \\
&=& \sum_{n=0}^\infty \left( (4-n) c_n^B(T) +
\bar{c}_n^B(T)\right) \left(\frac{\mu_B}{T}\right)^n \nonumber \\
              &\equiv& \sum_{n=0}^{\infty} s_n\left(\frac{\mu_B}{T}\right)^n.
\label{eq:es}
\end{eqnarray}
At present, we calculate the expansion coefficients $\bar{c}_n^B$ from
the coefficients $c_n^B$, in accordance with Eq.~\ref{eq:cne3p}, by
performing the $T$ derivative numerically, which introduces a small
systematic error.

In Fig.~\ref{fig:coff} we show the second, fourth and sixth order expansion
coefficients of the pressure, energy density and entropy density as given
in Eqs.~\ref{eq:PTaylor_hadronic} and \ref{eq:es}, obtained with the
p4fat3 action. Full symbols are from $N_\tau=4$ lattices, while
the open symbols denote results from $N_\tau=6$ lattices. We again find 
small cut-off effects, however, higher order derivatives of pressure, energy
density and entropy density with respect to $\mu_B$ are still very 
preliminary, as the error bars are large. This is especially true for 
the results from $N_\tau=6$ lattices. Nevertheless, the overall pattern of the
coefficients is in agreement with expectations based on an analysis of the singular
behavior of the free energy, making use of an appropriate scaling Ansatz.

\begin{figure*}
\begin{center}
\resizebox{0.49\textwidth}{!}{%
  \includegraphics{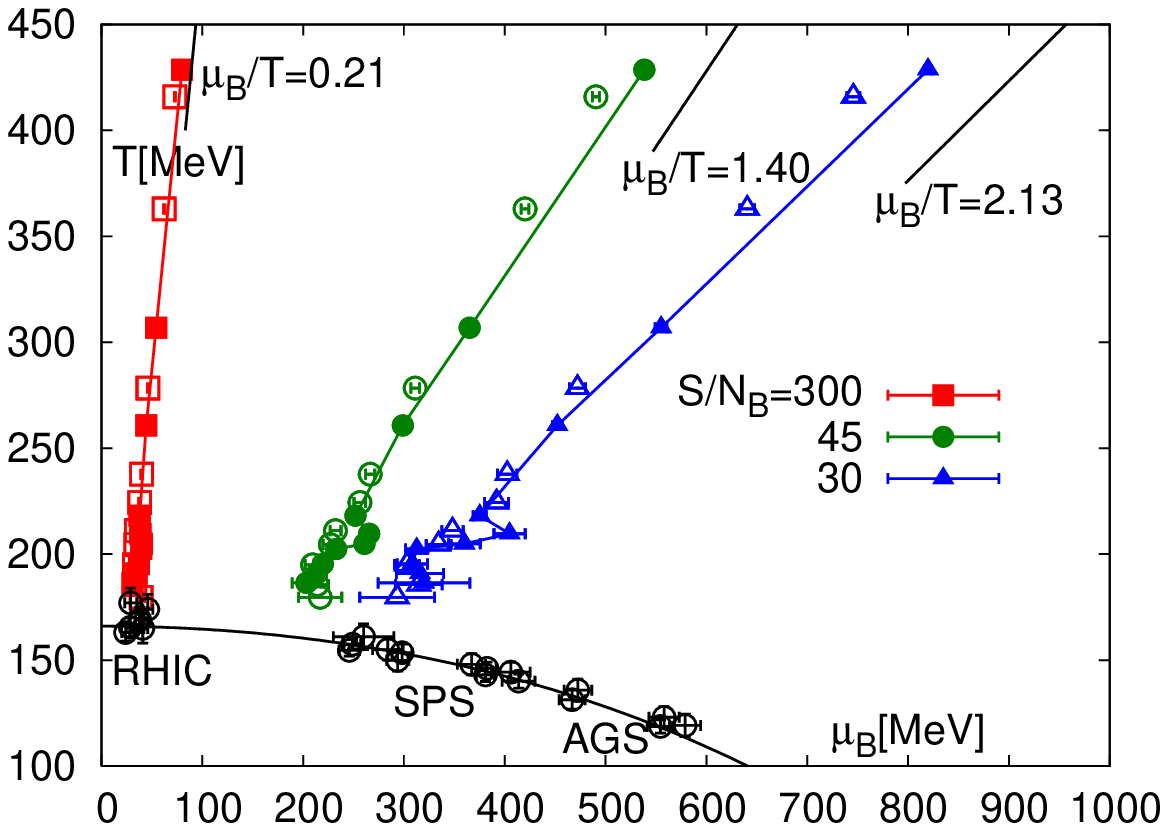}
}
\resizebox{0.49\textwidth}{!}{%
  \includegraphics{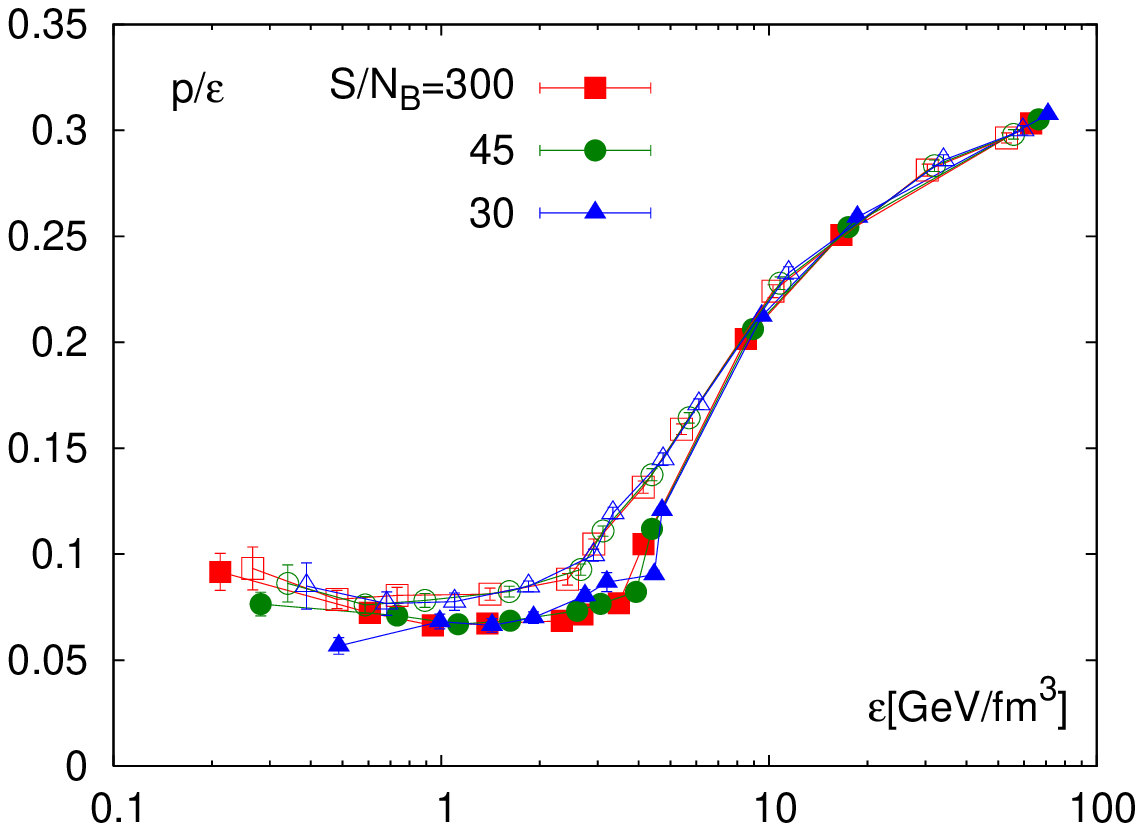}
}
\end{center}
\caption{On the left panel we show Isentropic trajectories in the
($T-\mu_B$)-diagram, corresponding to $s/n_B=300,45,30$ respectively.
Open symbols are form $N_\tau=6$, obtained by a 4th order Taylor
expansion of the pressure. Filled symbols are from $N_\tau=4$
calculations. We also show freeze-out data, as well as a
parameterization of the freeze-out curve from
\cite{redlich05}. Corresponding free gas limits are $\mu_B/T=0.21,
1.40, 2.13$ respectively and are indicated by solid lines.  On the
right panel we plot the the ratio of pressure and energy density along
those trajectories.}
\label{fig:isen_eos}
\end{figure*}

We find that the magnitude of the coefficients is decreasing drastically
with increasing order, for all analyzed temperatures. Thus an
approximation of the equation of state for small baryon chemical
potential by means of a fourth or sixth order expansion seems to be
justified. In general, an analysis of the radius of convergence of
such a Taylor series is of great interest for an analysis of the QCD
phase diagram, since the radius of convergence is bounded by the
location of the QCD critical point as well as by any first order phase
transition line.

\section{The isentropic equation of state}
\label{sec:isen}
By using the Taylor expansion coefficients of the baryon number
(Eq.~\ref{eq:nB_hadronic}) and entropy density (Eq.~\ref{eq:es}), we
can compute the ratio of entropy per baryon number as function of $T$
and $\mu_B$. Solving numerically for a constant ratio of entropy per
baryon number, $s/n_B$, we determine isentropic trajectories in the
$(T,\mu_B)$-plane.  These trajectories are relevant for the
description of matter created in relativistic heavy ion collisions.
After equilibration the dense medium created in such a collision will
expand along lines of constant entropy per baryon. It then is of
interest to calculate thermodynamic quantities along such isentropic
lines.

We find that isentropic expansion at high temperature is well
represented by lines of constant $\mu_B/T$ down to temperatures close
to the transition, $T\simeq 1.2 T_0$. In the low temperature regime we
observe a bending of the isentropic lines in accordance with the
expected asymptotic low temperature behavior. The isentropic expansion
lines for matter created at SPS correspond to $s/n_B \simeq 45$ while
the isentropes at RHIC correspond to $s/n_B \simeq 300$. The energy
range of the AGS which also corresponds to an energy range relevant
for future experiments at FAIR/Darmstadt is well described by $s/n_B
\simeq 30$. These lines are shown in Fig.~\ref{fig:isen_eos} (left)
together with data points characterizing the chemical freeze-out of hadrons
measured at AGS, SPS and RHIC energies.  These data points have been
obtained by comparing experimental results for yields of various
hadron species with hadron abundances in a resonance gas
\cite{cleymans,redlich05}. The solid curve shows a phenomenological
parameterization of these {\it freeze-out data} \cite{redlich05}.  In
general our findings for lines of constant $s/n_B$ are in good
agreement with phenomenological model calculations that are based on
combinations of ideal gas and resonance gas equations of state at high
and low temperature, respectively \cite{shuryak,toneev}.

Results shown in Fig.~\ref{fig:isen_eos} are based on a fourth order
expansion of the pressure. We find, however, that the truncation
error is small, i.e. the results change only little when we
consider also the sixth order term in $\mu_B$. In accordance with the 
good convergence of our results, we find, that all trajectories shown 
in Fig.~\ref{fig:isen_eos} (left)
are well within the radius of convergence of the Taylor series.
At present we estimate the radius of convergence of the pressure series
to $(\mu_B/T)^{\rm crit}\gsim 2.7$. The cut-off effects can be
estimated by comparing open and full symbols.

We now proceed and calculate energy density and pressure on lines of
constant entropy per baryon number using our Taylor expansion results
up to ${\cal O}(\mu_B^4)$. We find that both quantities obtain
corrections of about 10$\%$ at AGS (FAIR) energies ($s/n_B=30$) and
high temperatures.  The dependence of $\epsilon$ and $p$ on $s/n_B$
cancels to a large extent in the ratio $p/\epsilon$, which is most
relevant for the analysis of the hydrodynamic expansion of dense
matter. This may be seen by considering the leading ${\cal
O}(\mu_B^2)$ correction,
\begin{equation} \frac{p}{\epsilon} = \frac{1}{3} - \frac{1}{3}
\frac{\epsilon_0-3p_0}{\epsilon_0} \left( 1 + \left[
\frac{\bar{c}_2}{\epsilon_0-3p_0} - \frac{\epsilon_2}{\epsilon_0}
\right] \left( \frac{\mu_B}{T} \right)^2 \right) \; .
\label{povere}
\end{equation} In Fig.~\ref{fig:isen_eos} (right) we show $p/\epsilon$
as function energy density along our three isentropic trajectories. The
softest point of the equation of state is found to be
$(p/\epsilon)_{min} \simeq 0.07-0.09$,
for $N_\tau=4$ and $6$ respectively. Within our current numerical
accuracy it is independent of $s/n_B$. Similar results for the asqtad
action have been obtained in~\cite{milc_dens}. However, as our data is
preliminary, the analysis clearly suffers from poor statistics, which is 
in particular true for our $N_\tau=6$ results.

\section{Conclusions}
\label{sec:con}
We have presented results on the equation of state on lattices of 
$N_\tau=4,6$ \cite{EoS,milc_eos}
and $8$ \cite{hotQCDeos} with two different kinds of improved staggered 
fermions. Our
masses have been kept constant in physical units and are chosen such
that we have a physical strange quark mass ($m_s$) and 2 light quarks
with a mass of $m_l=0.1 m_s$. We also presented some preliminary results
with physical quark masses $m_l=0.05m_s$. We find that our two actions lead 
to a
consistent picture of the thermodynamics of QCD and find in particular
for the $N_\tau=8$ results only small cut-off effects. We have
calculated the equation of state as well as the velocity of sound and
find the softest point of the equation of state to be
$(p/\epsilon)_{min} \simeq 0.09$ at energy densities of $1-2$~GeV/fm$^3$.

Furthermore, we calculated corrections to the equation of state
arising from a non-zero baryon chemical potential, by means of a
Taylor expansion of the pressure. Within this framework we calculated
the isentropic equation of state along lines of constant entropy per
baryon number ($s/n_B$) for RHIC, SPS and AGS (FAIR) energies. Within
our current, preliminary, analysis we find the softest point of the
equation of state to be independent of $s/n_B$.

\section*{Acknowledgments}
This work has been supported in part by contracts DE - AC02 - 98CH10886
and DE - FG02 - 92ER40699 with the U.S. Department of Energy.
Numerical simulations have been performed on the BlueGene/L computers
at Laurence Livermore National Laboratory (LLNL) and the New York Center
for Computational Sciences (NYCCS) as well as on the QCDOC computer of 
the RIKEN-BNL research center, the DOE
funded QCDOC at Brookhaven National Laboratory (BNL) and the apeNEXT 
at Bielefeld University.

\end{document}